\documentclass[graybox]{SNmult}

\usepackage{type1cm}        %
\usepackage{makeidx}         %
\usepackage{graphicx}        %
\usepackage{multicol}        %
\usepackage[bottom]{footmisc}%

\usepackage{newtxtext}       %
\usepackage[varvw]{newtxmath}       %
\usepackage[caption=false]{subfig}

\makeindex             %

\begin{document}
\motto{Fast solvers for faster e-machine designs.}
\title*{Accelerating Industrial Finite Element Simulations of Electric Machines based on Runtime Analysis}
\titlerunning{Accelerating E-Machine Finite Element Simulations}
\author{Arvinth Shankar \and\\ Iryna Kulchytska-Ruchka\orcidID{0000-0002-9853-8006} \and\\Sebastian Schöps\orcidID{0000-0001-9150-0219}}
\authorrunning{Arvinth Shankar et al.}
\institute{Arvinth Shankar \at Robert Bosch GmbH, Robert-Bosch-Campus 1, 71272 Renningen, Germany, \email{arvinth.shankar@de.bosch.com} 
\and Iryna Kulchytska-Ruchka \at Robert Bosch GmbH, Robert-Bosch-Campus 1, 71272 Renningen, Germany, \email{iryna.kulchytska-ruchka@de.bosch.com}
\and Sebastian Schöps \at Computational Electromagnetics Group, Technische Universit\"at Darmstadt, Schloßgartenstraße 8, 64289 Darmstadt, Germany, \email{sebastian.schoeps@tu-darmstadt.de}
}
\maketitle
\abstract*{The simulation of electric-machines plays a significant role in the design of efficient and competitive products. Faster simulations reduce computational costs, such as CPU hours, and shorten development cycles, thereby enabling faster design iterations and ultimately accelerating time-to-market. In this work, we analyze the dominant computational bottlenecks and demonstrate how targeted acceleration measures can significantly reduce the overall runtime in industrially relevant 2D and 3D finite element simulations using an in-house code.
}

\abstract{The simulation of electric machines plays a significant role in the design of efficient and competitive products. Faster simulations reduce computational costs, such as CPU hours, and shorten development cycles, thereby enabling faster design iterations and ultimately accelerating time-to-market. In this work, we analyze the dominant computational bottlenecks and demonstrate how targeted acceleration measures can significantly reduce the overall runtime of 2D and 3D finite element simulations of electric machines in an industrial environment.
}

\section{Introduction}
\label{Bosch:SolverSteps}
Computationally efficient Finite Element (FE) simulations of electric machines depend on the performance of its various building blocks: mesher, finite element discretization, time stepping, (non-)linear solvers, post-processing etc. The required efficiency, however, depends on the use case. For high-fidelity simulations, both parallelization and algorithmic speed-up strategies can be effectively combined to reduce the wall-clock time. In contrast, when a large number of simulations are executed concurrently, parallelization resources are already utilized across runs. Thus the focus shifts to energy efficiency, where algorithmic speed-up is the key to reducing the computational cost per simulation. In practice, the overall runtime is driven by the repeated execution of several solver steps rather than any single operation. Here, we investigate the typical nonlinear solver steps: S$1$:~Residual evaluation, S$2$:~Jacobian evaluation, S$3$:~Line search (e.g. using Wolfe-Powell conditions), S$4$:~Matrix factorization or preconditioning, S$5$:~Linear system solution (direct or iterative).
Note that Wolfe-Powell conditions \cite{Nocedal:WolfePowell} include within the line search further residual evaluations in Armijo’s condition S$3a$ (used for sufficient decrease) and further Jacobian evaluations in the curvature condition S$3b$ (used for sufficient change in gradient). This implies that the computational cost of residual
evaluation S$1$ or Jacobian evaluation S$2$ also affect the cost of
the line search procedure S$3$, since the line search may require multiple such evaluations within its iterative procedure. In the following sections, we analyze the computational time spent in each solver step for representative 2D and 3D simulations of e-machines, identify the main bottlenecks, and describe the acceleration measures implemented to reduce their contribution to the overall runtime.

\section{A 2D static case}
\label{Bosch:2dStaticCaseSection}
\begin{figure}[!htbp]
	\subfloat[2D PMSM \label{Bosch:2dPMSM}]{\includegraphics[width=.45\textwidth]{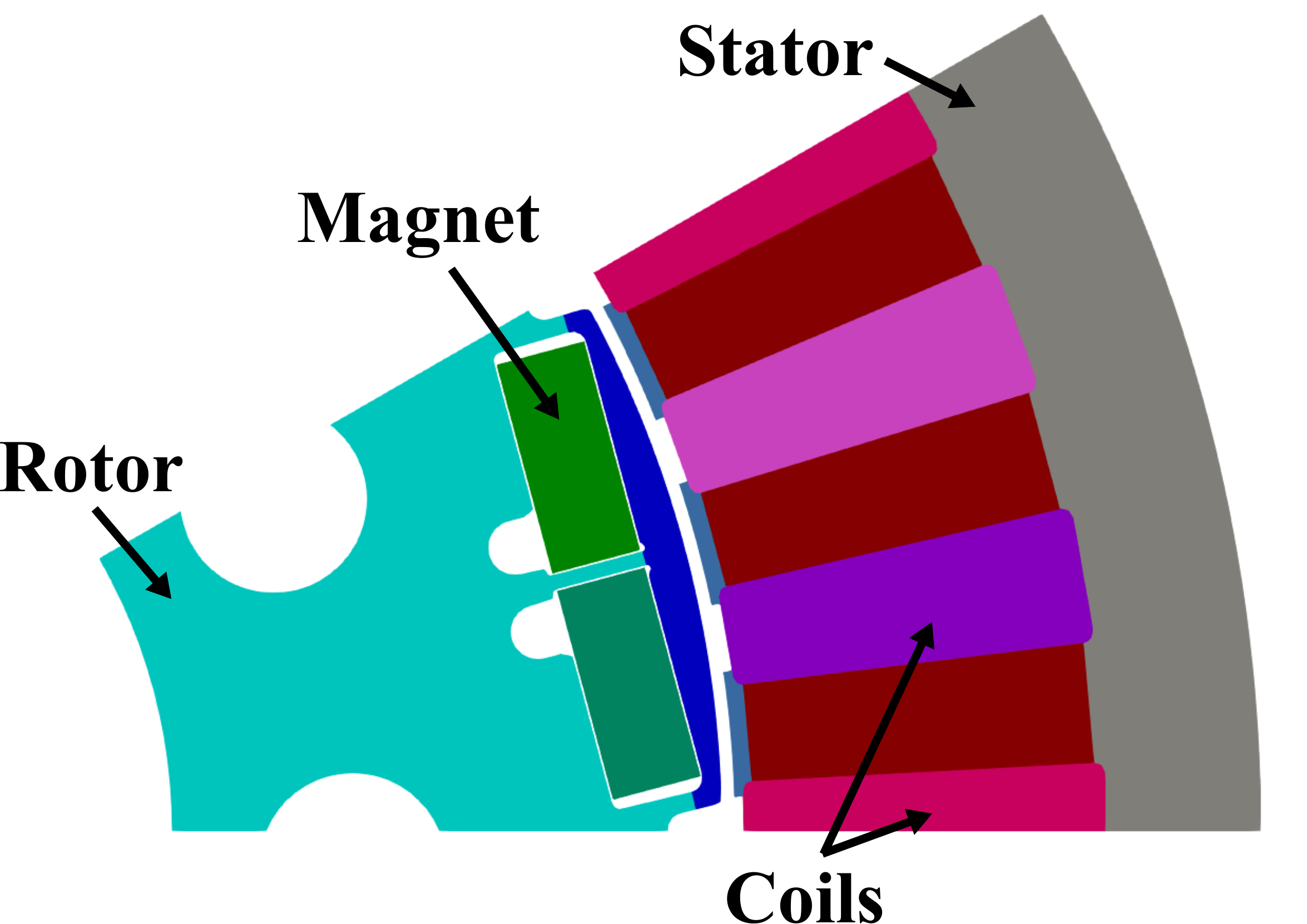}}
	\quad
	\subfloat[Nonlinear $B$-$H$ curve using a spline approximation for the material M330-35A \label{Bosch:BHcurve}]{\includegraphics[width=.47\textwidth]{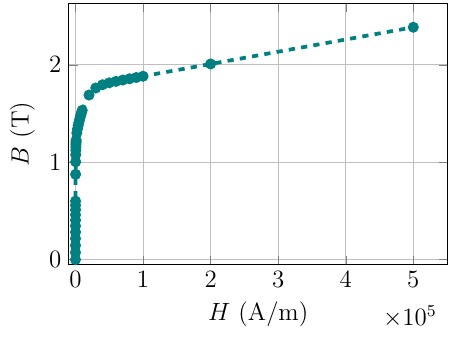}}
	\caption{A 2D model and $B$-$H$ curve}
	\label{Bosch:2DmodelAndBHcurve}
\end{figure}

\begin{figure}[!htbp]
	\sidecaption[t]
	\centering
	\includegraphics[width=5cm]{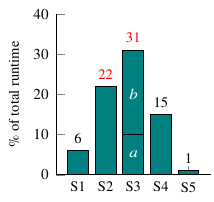}
	\caption{Contribution to the total runtime of the principal solver steps. S$1$:~Residual evaluation: S$2$:~Jacobian evaluation, S$3$:~Line search (e.g. using S$3a$: Armijo condition and S$3b$: curvature condition), S$4$:~Matrix factorization, S$5$:~Linear system solution (direct)}
	\label{Bosch:2dModelAndBottleneck}
\end{figure}

We consider a 2D model of Permanent-Magnet Synchronous Machine (PMSM), shown in Fig.~\ref{Bosch:2dPMSM}, discretized with $10{,}000$ elements. Both the stator and rotor are modeled using materials characterized by the nonlinear $B$-$H$ curves shown in Fig.~\ref{Bosch:BHcurve}. The stator coils are excited with $3$-phase sinusoidal currents and a magnetostatic simulation was performed for $90$ rotor positions.
Figure~\ref{Bosch:2dModelAndBottleneck} shows the percentage of total runtime spent in each solver step, as defined in Sect.~\ref{Bosch:SolverSteps}. 
We observe that the most time-consuming steps are S$2$ and S$3$, namely, the Jacobian evaluation required by the Newton solver and the line-search sub-iterations. The model was also simulated on meshes with different densities, namely, using $3{,}000$ and $40{,}000$ elements, and we obtained nearly the same relative time contributions as in Fig.~\ref{Bosch:2dModelAndBottleneck}. Additionally, for a fixed mesh, the effect of varying angular step sizes and Newton tolerances was examined. In all cases, the relative runtime contributions showed no significant variation, confirming that the identified bottlenecks are consistently reproduced. Based on these bottlenecks, our targeted speed-up measures include the use of Armijo condition only in the line search step S$3$ and the parallelization of (bi)linear form creation among FEs using multi-threading. Each of these measures is discussed in a separate section.
\subsection{Newton with Armijo-only line search}
\label{Bosch:NewtonWithArmijo-onlyLineSearch}
Consider the nonlinear system of equations
\begin{equation}
	\vec{F}(\vec{a}^i) = \left(\mathbf{K}_{\boldsymbol{\nu}}(\vec{a}^i) +   \frac{\mathbf{M}}{\Delta t^i} \right)\vec{a}^i - \vec{j}^i - \frac{\mathbf{M}}{\Delta t^i} 
	\,\vec{a}^{i-1} = \vec{0}\;,
	\label{Bosch:NonlinearSystemEqns}
\end{equation}
where $\vec{F}$ denotes the nonlinear residual function, $\vec{a}^i$ is the unknown magnetic vector potential at time step $i$, and $\Delta t^i$ is the corresponding time-step size. Furthermore, $\mathbf{K}_{\boldsymbol{\nu}}$ is the nonlinear curl-curl reluctivity matrix, $\mathbf{M}$ denotes the (singular) conductivity matrix, which does not appear in static formulations and $\vec{j}$ is the prescribed source current-density vector. Equation~(\ref{Bosch:NonlinearSystemEqns}) arises from the eddy-current approximation of Maxwell's equations in the modified magnetic vector potential formulation \cite{Kuczmann:FEmagnetics} followed by FE spatial and (implicit Euler) time discretization. Reluctivity matrix $\mathbf{K}_{\boldsymbol{\nu}}$ in (\ref{Bosch:NonlinearSystemEqns}) is defined by the bilinear form over the computational domain~$\Omega$ as
\begin{equation}
	(\mathbf{K}_{\boldsymbol{\nu}}(\vec{a}))_{ij} = \int\limits_\Omega \textnormal{curl}\,\vec{w}_i\cdot \boldsymbol{\nu}\,\,\textnormal{curl}\,\vec{w}_j\,d\Omega \;.
\end{equation}
Here $\boldsymbol{\nu}$ is the reluctivity, $\vec{w}_i$ and $\vec{w}_j$ are the basis functions of the edge FE space, where $i$ and $j$ denote the row and column indices of $\mathbf{K}_{\boldsymbol{\nu}}$, respectively. Similarly, the conductivity matrix $\mathbf{M}$ in (\ref{Bosch:NonlinearSystemEqns}) is defined by
\begin{equation}
	\mathbf{M} = \int\limits_\Omega \vec{w}_i\cdot \sigma\,\vec{w}_j\,d\Omega\;,
\end{equation}
where $\sigma$ is the electrical conductivity.
A classical approach for solving (\ref{Bosch:NonlinearSystemEqns}) is Newton's method \cite{Deuflhard:NewtonMethod}, in which the nonlinear system is linearized and Newton iterations are performed at each time step until convergence is reached. At Newton iteration $k$, the solution vector is updated as
$\vec{a}_{k+1} = \vec{a}_{k} + \Delta\vec{a}_k$, where $k$ denotes the iteration number and $\Delta\vec{a}_k$ is the Newton search direction obtained from the solution of linearized system at the current time step as
\begin{equation}
	\vec{F^{\prime}}(\vec{a}_k)\Delta\vec{a}_k = -\vec{F}(\vec{a}_k)\;.
	\label{Bosch:stepDirectionFormula}
\end{equation}
Here $\vec{F^{\prime}}(\vec{a}_k)$ represents the Jacobian evaluated at the current solution iterate~$\vec{a}_k$. A more robust approach is to combine Newton's method with a line search strategy~\cite[pp.~30--35]{Nocedal:WolfePowell}, which selects an appropriate step length $\alpha_k$ along the search direction to ensure adequate reduction in residual function. Accordingly, the solution update becomes
\begin{equation}
	\vec{a}_{k+1} = \vec{a}_{k} + \alpha_k\Delta\vec{a}_k\;, \quad \alpha_k \in (0,1]\;.
\end{equation}
The basic idea of a line-search procedure is to define the objective function $\psi$ as
\begin{equation}
	\psi(\vec{a}_{k}) = \frac{1}{2}\left(\vec{F}(\vec{a}_{k})\right)^{\top}\cdot \vec{F}(\vec{a}_{k}) = \frac{\left\|\vec{F}(\vec{a}_{k})\right\|^{2}}{2}\;,
	\label{Bosch:LineSearchObjective}
\end{equation}
and to determine the step length $\alpha_k$ such that the objective function decreases sufficiently. 
A practical approach for determining the step length is through an inexact line search condition, i.e. Armijo condition (S$3a$ in Fig.~\ref{Bosch:2dModelAndBottleneck}) of the form
\begin{equation}
	\psi(\vec{a}_{k} + \alpha_k\Delta\vec{a}_k) \leq \psi(\vec{a}_{k}) + \sigma \alpha_k \nabla \psi(\vec{a}_{k})^\top \Delta\vec{a}_k \eqcolon g(\alpha_k)\;,
	\label{Bosch:Armijo}
\end{equation}
\begin{figure}[!htbp]
	\subfloat[Function decrease condition \label{Bosch:sufficientFuncDecrease}]{\includegraphics[width=.485\textwidth]{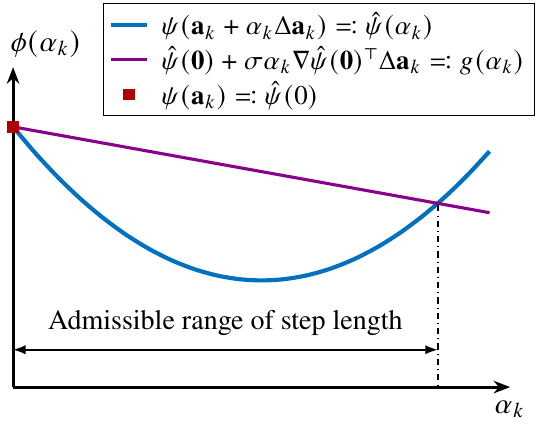}}
	\quad
	\subfloat[Curvature condition
	\label{Bosch:sufficientSlopeChange}]{\includegraphics[width=.485\textwidth]{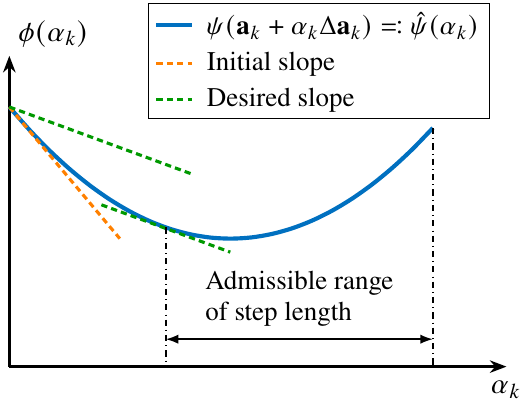}}
	\caption{Acceptable range of step length $\alpha_k$ using line search conditions}
	\label{Bosch:AcceptableStepLengths}
\end{figure}
where $\sigma \in (0,1)$ is referred to as the Armijo parameter and is typically chosen to be a small value, e.g. $10^{-4}$. The inequality condition~(\ref{Bosch:Armijo}) defines the admissible interval of step lengths~$\alpha_k$ as illustrated in Fig.~\ref{Bosch:sufficientFuncDecrease}.
This is complemented by a curvature condition~(S$3b$ in Fig.~\ref{Bosch:2dModelAndBottleneck}), that enforces a sufficient change in gradient as
\begin{equation}
	\nabla \psi(\vec{a}_{k} + \alpha_k \Delta\vec{a}_k)^\top \Delta\vec{a}_k \geq \rho\, \nabla \psi(\vec{a}_{k})^\top \Delta\vec{a}_k\;.
	\label{Bosch:Curvature}
\end{equation}
Here $\rho \in (\sigma, 1)$, where $\sigma$ is the parameter from (\ref{Bosch:Armijo}). This condition requires the slope of the objective function at the trial point to be greater than $\rho$ times the initial slope. This implies that a strongly negative slope indicates that the objective function is decreasing along the search direction, whereas a slightly negative or positive slope suggests that no further decrease of the function is expected in the chosen direction. This behavior is illustrated in Fig.~\ref{Bosch:sufficientSlopeChange}.
Equation~(\ref{Bosch:Armijo}) together with (\ref{Bosch:Curvature}) constitutes the Wolfe-Powell conditions.

The line search S$3$ shown in Fig.~\ref{Bosch:2dModelAndBottleneck} is performed based on the inequality conditions~(\ref{Bosch:Armijo}) and (\ref{Bosch:Curvature}), and contributes $31\%$ of the total runtime. It was observed that nearly two-thirds of the line-search runtime come from repeated evaluation of the Jacobian $\mathbf{F^{\prime}}$ associated with the curvature condition (\ref{Bosch:Curvature}) within the line-search sub-iterations. To reduce this computational cost, the line-search is restricted to consider only the Armijo condition (\ref{Bosch:Armijo}). The step length is then updated by backtracking until the condition is satisfied, using the backtracking parameter~$\beta$ as follows
\begin{equation}
	\alpha_k^{\,j+1} = \beta\,\alpha_k^{\,j}\,, \quad \beta = 0.5\,.
	\label{Bosch:backtrackingArmijo}
\end{equation}
Here, $\alpha_k^{\,j}$ represents the step length from the previous backtracking iteration~$j$ within the line-search algorithm. The parameter $\beta$ controls the rate at which the trial step length is reduced during the backtracking procedure. A lower bound is also imposed on the step length to prevent it from becoming excessively small during backtracking. From a computational performance perspective, this modification avoids expensive repeated Jacobian evaluations inside the line-search routine, thereby making the line-search procedure approximately three times faster. Applying this speed-up measure to the simulation discussed in Sect.~\ref{Bosch:2dStaticCaseSection} resulted in an overall runtime reduction of approximately $25\%$, while retaining the accuracy. A further speed-up strategy may also be explored through a self-adaptive initial step-length scheme, which is discussed in the following section.
\subsubsection{Self-adaptive initial step length}
\label{Bosch:Self-adaptiveInitialStepLength}
The original line-search procedure determines a suitable step length starting from an initial value for the step length $\alpha_k^{0}$. This initial guess is typically chosen as the maximum admissible value, $\alpha_k^{0} = 1$. The line-search procedure is then initialized with this value at each Newton iteration. 

Depending on the degree of nonlinearity of the problem, the accepted step-length may deviate substantially from the initial guess. In such cases, the line-search sub-iterations may require multiple evaluations of the residual and Jacobian to satisfy the Armijo (\ref{Bosch:Armijo}) and curvature condition (\ref{Bosch:Curvature}), which might lead to high computational cost. Numerical observations nevertheless indicate that, for most iterations, the accepted step lengths at successive Newton iterations remain close to one another.

A self-adaptive initial step-length scheme based on a nonlinear line-search formulation, i.e. nonlinear dependency of line-search criterion on the step length, for traffic assignment algorithms is described in~\cite{Chen:Self-adaptiveArmijo}. In the present study, a similar self-adaptive update strategy with linearized sub-problem using the standard Armijo-type condition is formulated. The initial value is adapted using the step length accepted in the preceding Newton iteration. Importantly, this initial value is allowed to both increase and decrease, depending on the outcome of the previous iteration. Such a non-monotone update strategy allows the method to recover full step length $\alpha_k = 1$ when the iterate is sufficiently close to the true solution, thereby enabling fast local quadratic convergence  of Newton's method, while still permitting smaller step lengths in regions of strong nonlinearity. The resulting self-adaptive step-length procedure is described below.

\begin{figure}[!h]
	\centering
	\sidecaption[b]
	\includegraphics[width=7cm]{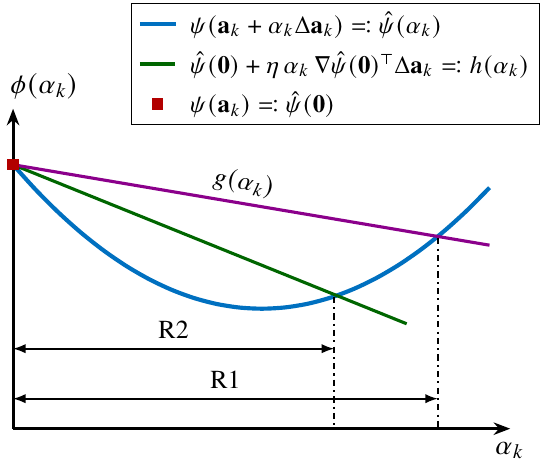}
	\caption{Admissible step-length ranges, where R$1$ denotes the admissible range based on the original Armijo condition~(\ref{Bosch:Armijo}) and R$2$ denotes the admissible range based on a stricter condition~(\ref{Bosch:selfAdaptiveArmijo}) for adaptivity.}
	\label{Bosch:selfAdaptiveStepSize}
\end{figure}

Let $\alpha_k^\ast$ denote the accepted step length at Newton iteration $k$, and let $\bar{\alpha}\textsubscript{max}$ denote the upper bound on the step length. In our case, $\bar{\alpha}\textsubscript{max} = 1$. At the end of Newton iteration $k$, the initial step length for the line search in Newton iteration $k+1$ is updated adaptively based on the accepted $\alpha_k^\ast$ through the following condition,
\begin{equation}
	\psi(\vec{a}_{k} + \alpha_k^\ast\Delta\vec{a}_k) \leq \psi(\vec{a}_{k}) + \eta\,\alpha_k^\ast\, \nabla \psi(\vec{a}_{k})^\top \Delta\vec{a}_k \eqcolon h(\alpha_k^\ast)\;.
	\label{Bosch:selfAdaptiveArmijo}
\end{equation}
This condition is the same as the Armijo condition in (\ref{Bosch:Armijo}), except that the parameter~$\eta$ is used instead of $\sigma$. Here $\eta \in (\sigma,1]$. In this study, $\eta$ is chosen as a relatively large value, e.g. $\eta = 0.5$, to check whether a further residual decrease is obtained when using the accepted $\alpha_k^\ast$. The admissible step lengths defined by this condition are illustrated in Fig.~\ref{Bosch:selfAdaptiveStepSize}.
If condition~(\ref{Bosch:selfAdaptiveArmijo}) is satisfied, then the initial step length for the next Newton iteration, i.e. $\alpha_{k+1}^0 $ is updated as
\begin{equation}
	\alpha_{k+1}^0 = \min\left(\tau\alpha_k^\ast\,, \bar{\alpha}\textsubscript{max}\right)\;.
	\label{Bosch:updateSteplength}
\end{equation}
Here, $\tau > 1$ is a constant. The update in~(\ref{Bosch:updateSteplength}) is based on the observation that if the accepted step length from the previous iteration already satisfies~(\ref{Bosch:selfAdaptiveArmijo}), the resulting residual reduction is sufficient, allowing the initial step length for the next iteration to be increased. The upper bound $\bar{\alpha}_{\max}$ prevents the step length from becoming too large and causing overshooting. In this study, $\tau=2$ is chosen so that the previously accepted step length is doubled when needed, complementing the backtracking factor $\beta=0.5$ in (\ref{Bosch:backtrackingArmijo}), which reduces the step length by half. This allows the method to return quickly to the unit step length. If~(\ref{Bosch:selfAdaptiveArmijo}) is not satisfied, the initial step length for the next iteration is kept equal to the accepted step length from the preceding iteration, i.e.~$\alpha_{k+1}^0 = \alpha_k^\ast$.

To assess the proposed self-adaptive initialization strategy, the number of residual evaluations per Newton iteration is compared with that of a fixed initial step length for the simulation setup in Sect.~\ref{Bosch:2dStaticCaseSection}.
A difference is observed at the first rotor position, where the simulation is initialized from a zero solution.
\begin{figure}[!h]
	\centering
	\sidecaption[b]
	\includegraphics[width=7.7cm]{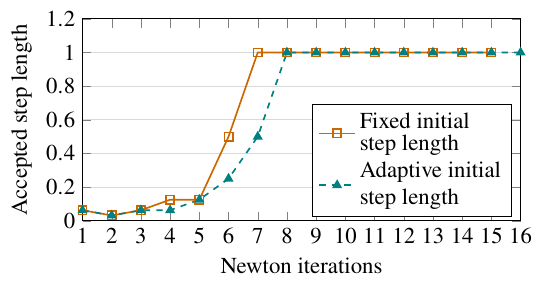}
	\caption{Comparison of accepted step lengths per Newton iteration at the first rotor position using fixed and adaptive initial step-length strategies}
	\label{Bosch:StepLengthvsNewton}
\end{figure}
\begin{figure}[!h]
	\centering
	\sidecaption[b]
	\includegraphics[width=7.7cm]{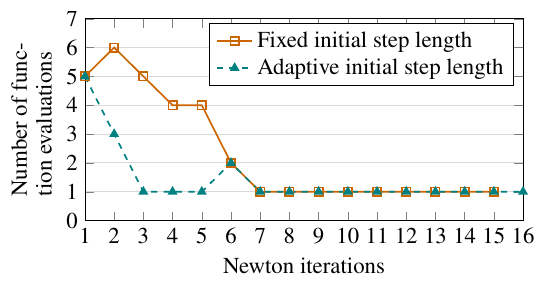}
	\caption{Comparison of line-search residual evaluations per Newton iteration at the first rotor position using fixed and adaptive initial step-length strategies}
	\label{Bosch:FuncEvalvsNewton}
\end{figure}
As shown in Fig.~\ref{Bosch:StepLengthvsNewton}, the accepted step lengths $\alpha_k^\ast$ in the initial iterations are nearly identical for both strategies. However, a noticeable difference in the number of residual evaluations occurs in the first few Newton iterations as shown in Fig.~\ref{Bosch:FuncEvalvsNewton}. The adaptive strategy required a total of $23$ residual evaluations, compared to~$35$ for the fixed initialization, reflecting a clear reduction. The unit step length is recovered at the $8\textsuperscript{th}$ Newton iteration for the adaptive strategy versus the $7\textsuperscript{th}$ for the fixed scheme. The increase factor $\tau$ in~(\ref{Bosch:updateSteplength}) ensures this recovery within a single iteration, so that the savings in residual evaluations outweigh the cost of the additional Newton iteration required for the adaptive scheme. 

For this magnetostatic case, continuation from the preceding rotor position provides a well-conditioned initial guess at each subsequent position, so the unit step length $\alpha_k^\ast=1$ gets accepted without line-search backtracking. As a result, both strategies performed similarly across majority of rotor positions, indicating that the self-adaptive update introduces no overhead when the full Newton step, i.e. $\alpha_k^\ast=1$ is admissible.
The speed-up is therefore observed primarily at the first rotor position and yields only $1\%$ for the model with $10{,}000$ elements. However, if Newton iterations at each rotor position are started from a fixed (zero) field, e.g. when computing all rotor positions in parallel, the self-adaptive strategy yields a speed-up of approximately $3\%$ for $10{,}000$ elements and $7\%$ for the model with $40{,}000$ elements. Larger models may profit even more.
\subsection{Parallelization of (bi)linear forms}
\label{Bosch:ParallelizationOfBilinearForm}
For nonlinear problems such as (\ref{Bosch:NonlinearSystemEqns}), the Jacobian in (\ref{Bosch:stepDirectionFormula}) is required for the Newton iterations and depending on the line-search strategy, may also be required for the line-search procedure. The Jacobian matrix $\vec{F^{\prime}}(\vec{a})$ of the residual $\vec{F}(\vec{a})$, defined in~(\ref{Bosch:NonlinearSystemEqns}), at a given time step with step size $\Delta t$ is given by
\begin{equation}
	\vec{F^{\prime}}(\vec{a}) = \mathbf{K}_{\boldsymbol{\nu}_d} (\vec{a}) + 
	\frac{\mathbf{M}}{\Delta t}\;,
	\label{Bosch:ParallelJacobian}
\end{equation}
where $\boldsymbol{\nu}_d$ denotes the differential reluctivity, see e.g. \cite{Gersem:DiffMatMatrix} and the nonlinear matrix $\mathbf{K}_{\boldsymbol{\nu}_d}$ is obtained from the reluctivity matrix $\mathbf{K}_{\boldsymbol{\nu}}$ in~(\ref{Bosch:NonlinearSystemEqns}) as 
\begin{equation}
	\mathbf{K}_{\boldsymbol{\nu}_d} =
	\frac{\D}{\D \vec{a}}
	\left(\mathbf{K}_{\boldsymbol{\nu}}(\vec{a}) \vec{a}\right)\;.
\end{equation}
The Jacobian  $\vec{F^{\prime}}$ in (\ref{Bosch:ParallelJacobian}) and the residual $\vec{F}$ are assembled from individual element contributions. Since these element-level computations are local and mutually independent, additional speed-ups are achieved through multithreaded parallelization of the local matrix and vector evaluations required for the global assembly of the Jacobian and the residual. 
Applying this spatial parallelization to the Jacobian and residual computations for the simulation discussed in 
\begin{figure}[!htbp]
	\includegraphics[width=12cm]{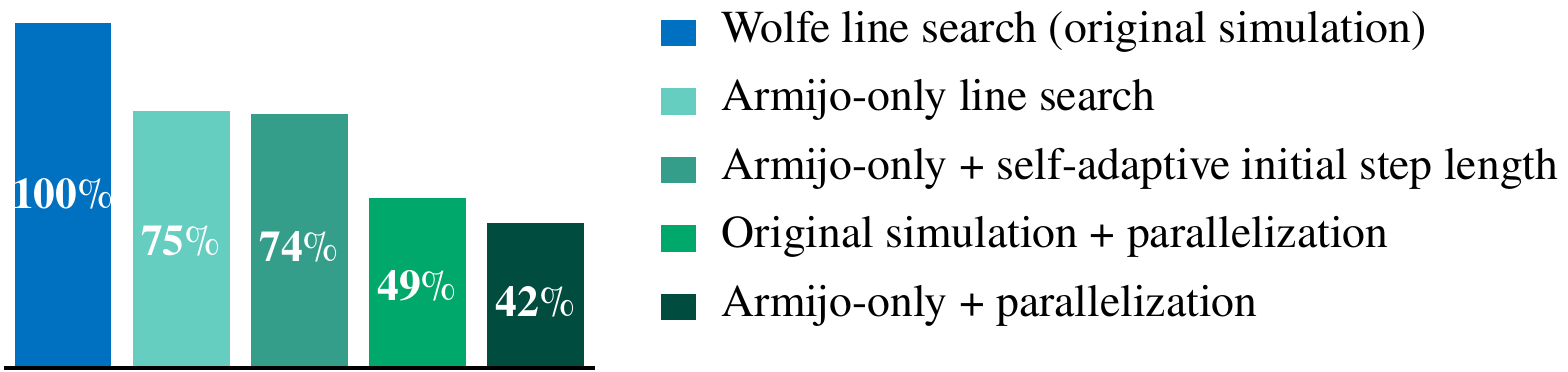}
	\caption{Relative simulation runtime using each speed-up measure, referenced to the original simulation}
	\label{Bosch:2dImprovedMeasures}
\end{figure}
Sect.~\ref{Bosch:2dStaticCaseSection} reduces the overall simulation time by approximately $30\%$ and $51\%$ using $8$ and $16$ threads, respectively. When this speed-up measure is combined with the strategy from Sect.~\ref{Bosch:NewtonWithArmijo-onlyLineSearch}, i.e. using only the Armijo condition with backtracking together with the parallelization, the total simulation time is reduced by approximately $46\%$ and $58\%$ using $8$ and $16$ threads, respectively.
Figure~\ref{Bosch:2dImprovedMeasures} shows the runtime improvements achieved by applying each acceleration measure individually and in combination, where the parallelization is performed using $16$ threads.
\section{A 3D transient case}
\label{Bosch:3DTransientCase}
\begin{figure}[!htbp]
	\subfloat[3D PMSM \label{Bosch:3dPMSM}]{\includegraphics[width=.42\textwidth]{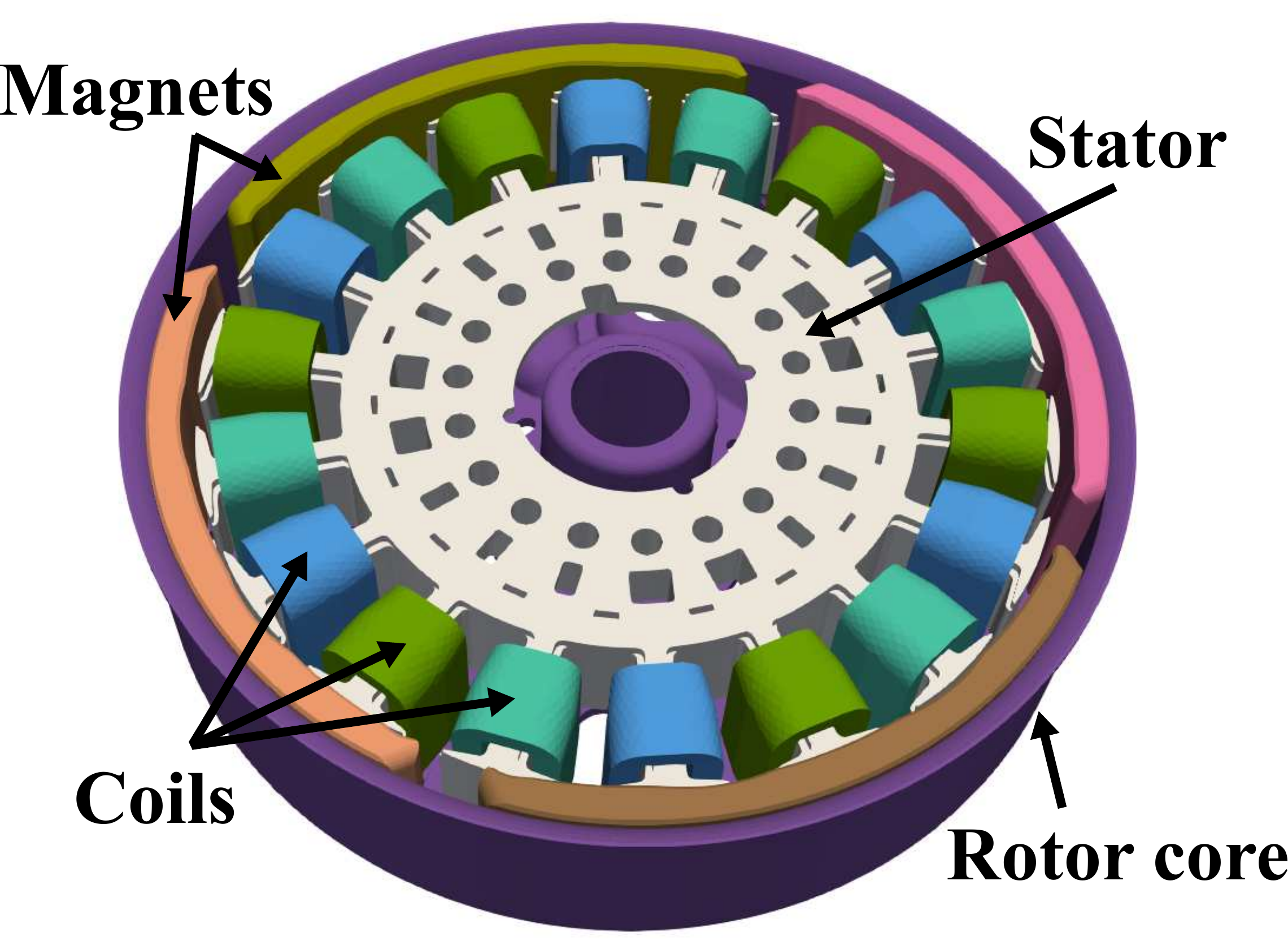}}
	\quad
	\subfloat[Runtime per solver step
	\label{Bosch:3DmodelAndSolverSteps}]{\includegraphics[width=.4\textwidth]{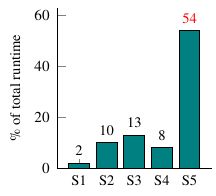}}
	\caption{3D model and contribution to the total runtime of the principal solver steps: S$1$:~Residual evaluation, S$2$:~Jacobian evaluation, S$3$:~Line search procedure, S$4$:~Preconditioning, S$5$:~Linear system solution (iterative)}
	\label{Bosch:3DModelAndBottlenecks}
\end{figure}

We now consider a 3D model of a PMSM, shown in Fig.~\ref{Bosch:3dPMSM} and meshed using $6$~million elements. The rotor magnets are assigned a nonzero electrical conductivity of $7 \times 10^4$~S/m and the stator coils are excited with~$3$-phase sinusoidal currents. An eddy current simulation is performed for $21$ time steps with a time-step size of $2.08 \times 10^{-4}$\,s, such that the rotor rotates by $1^\circ$ at each time step. 
Fig.~\ref{Bosch:3DmodelAndSolverSteps} shows that the bottleneck is the linear solver step~S$5$.
The simulation of 3D problems typically requires an iterative linear solver, since direct solvers are computationally and memory-inefficient, particularly as the problem size increases~\cite{Saad:Iterative}. 
Iterative linear solvers require adequate preconditioning techniques to ensure faster convergence. 
However, the efficiency of a preconditioner depends on the specific problem type. Classical Algebraic Multigrid (AMG) is one of the most widely adopted preconditioners \cite{Boyle:AMGPrecond}, which operates by correcting errors across different spatial grid levels that are constructed algebraically, and has been found to be particularly efficient for Poisson problems.

Since the Maxwell equations possesses a large gradient nullspace, a straightforward application of AMG is not efficient. A more established and computationally efficient preconditioner for problems in $\boldsymbol{H}(\textnormal{curl})$ space is the Auxiliary-Space Maxwell (AMS) preconditioner, developed by Hiptmair and Xu \cite{Hiptmair:AMS,Kolev:AMS}. This method decomposes the original $\boldsymbol{H}(\textnormal{curl})$ space, discretized by the N\'ed\'elec finite element space $\boldsymbol{V}_h$ into auxiliary nodal spaces and applies AMG within those auxiliary spaces, where it has been demonstrated to be particularly powerful.

Let $S_h$ and $\boldsymbol{S}_h$ denote the scalar $H^1$--conforming and vector $\boldsymbol{H}^1$--conforming FE spaces, respectively. The AMS method approximates the inverse of the original system matrix as
\begin{equation}
	\mathbf{P}^{-1} = \mathbf{R}^{-1} + \mathbf{G}_h \mathbf{P}_{\mathbf{G},h}^{-1} \mathbf{G}_h^\top + \mathbf{\Pi}_h \mathbf{P}_{\mathbf{\Pi},h}^{-1} \mathbf{\Pi}_h^\top\;,
	\label{Bosch:precond}
\end{equation}
where $\mathbf{P}$ is the resulting preconditioner, and $\mathbf{R}$ is the Gauss--Siedel smoother for the original system matrix. The operator $\mathbf{G}_h : S_h \to  \boldsymbol{V}_h$ defines a discrete gradient, and $\mathbf{\Pi}_h: \boldsymbol{S}_h\to\boldsymbol{V}_h$ is the interpolation operator. $\mathbf{P}_{\mathbf{G},h}$ and $\mathbf{P}_{\mathbf{\Pi},h}$ are AMG preconditioners, see e.g.~\cite{Kolev:AMS}.

For magnetostatic problems, gradient fields lie in the kernel of the system matrix due to the absence of the eddy-current term. Consequently, no gradient-space correction is required and the term $\mathbf{G}_h \mathbf{P}_{\mathbf{G},h}^{-1} \mathbf{G}_h^\top$ can be omitted. In contrast, for eddy-current problems, gradient correction becomes necessary. For this, Hypre~\cite{hypre:LawrenceLivermore} supports a variational construction of the nodal auxiliary matrix directly from the system matrix and the discrete gradient operator $\mathbf{G}_h$. However, we obtained improved Newton convergence by supplying an explicit auxiliary matrix, 
which the AMG solver uses directly to incorporate the required gradient correction term.
\begin{figure}[!htbp]
	\sidecaption[b]
	\centering
	\includegraphics[width=7.5cm]{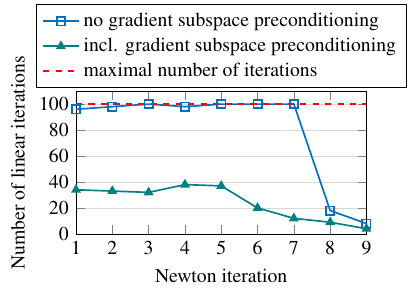}
	\caption{Influence of the gradient subspace preconditioning on the convergence of the linear solver at each Newton iteration of the first time step}
	\label{Bosch:LinearSolverConvergence}
\end{figure}
This auxiliary matrix is given by the Poisson operator, $(\beta\,\nabla \phi\,,\nabla \psi)$, where $\beta$ denotes the coefficient of the conductivity matrix term, and $\phi$, $\psi$ are the linear nodal basis functions, with $\phi, \psi \in \textit{S}_h$. The coefficient $\beta = \sigma / \Delta t$, where $\sigma$ is the electrical conductivity in the respective domain and $\Delta t$ is the time-step size.

Let us take a closer look at the influence of the gradient correction in AMS on the convergence of the linear solver for the 3D machine model (from Fig.~\ref{Bosch:3dPMSM}). The linear solver we use is the Minimum Residual (MINRES) method from PETSc~\cite{Petsc:MINRES}. The number of linear solver iterations required at each Newton iteration during the first time step is shown in Fig.~\ref{Bosch:LinearSolverConvergence}. The iterative solver required up to three times fewer iterations when the gradient subspace correction was applied, whereas without it the iteration limit of $100$ was often reached. This improvement is particularly pronounced during the initial time steps and reduces the total simulation time by $14\%$. 
\begin{figure}[!htbp]
	\includegraphics[width=11cm]{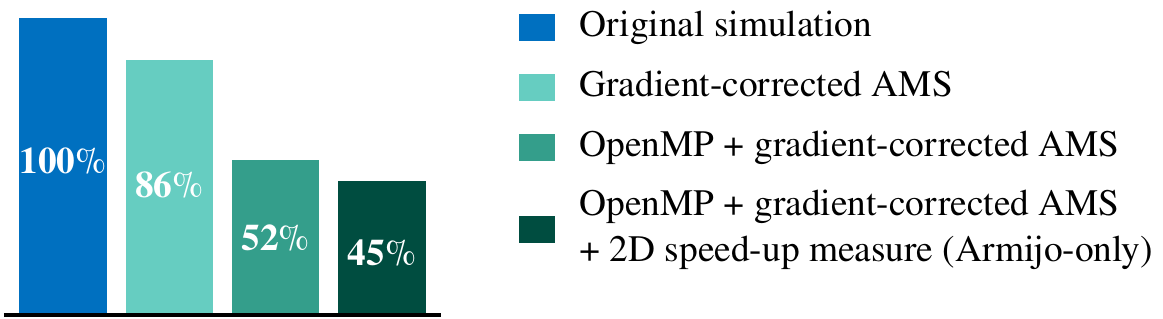}
	\caption{Relative simulation runtime using each speed-up measure, referenced to the original simulation}
	\label{Bosch:3dImprovedMeasures}
\end{figure}
Additional speed-ups of the preconditioner computation and the linear solver are achieved by compiling the algebraic libraries, namely Hypre and PETSc, with OpenMP multithreading.
OpenMP parallelization with 8 threads yield three times faster AMS preconditioning and two times faster linear system solution with MINRES. Combining the two acceleration measures for the 3d model (Fig.~\ref{Bosch:3dPMSM}), namely the gradient-corrected preconditioner and OpenMP parallelization (with $8$ threads), resulted in an overall runtime reduction of $48\%$, as shown in Fig.~\ref{Bosch:3dImprovedMeasures}. Further including the Armijo-only line search yields a total runtime reduction of $55\%$.
\section{Conclusion}
In this work, we identified the time-dominant solver steps and analyzed their relative contribution to the overall runtime of 2D and 3D FE e-machine simulations. For the 2D example, the Jacobian computation for Newton's method and the line-search procedure are identified as the most time-consuming solver components. These are accelerated through targeted measures including the use of Armijo-only condition with backtracking in the line search and a parallelized (bi)linear computation. Individually, these measures yield overall runtime reductions of $25\%$ and $51\%$, respectively, and when combined, overall runtime reduction of $58\%$ is achieved. For the 3D e-machine example, the linear iterative solver was identified as the dominant bottleneck, and the use of appropriate preconditioning results in a substantial reduction in the number of linear solver iterations, yielding an overall runtime reduction of $14\%$. In combination with OpenMP multithreading, an overall runtime reduction of $48\%$ is achieved using~$8$~threads. Further incorporating the 2D acceleration measure (Armijo-only) into the 3D example yields an overall runtime reduction of $55\%$.

\end{document}